\documentstyle[12pt,aaspp4]{article}
 
\newcommand{\etal}{{\it et al.}\ }

\lefthead{von Hippel, Bothun, \& Schommer}
\righthead{The WDMF and Supernova Ia Luminosities}

\begin{document}

\title{Stellar Populations and the White Dwarf Mass Function:\\
Connections To Supernova Ia Luminosities}

\author{Ted von Hippel}
\affil{Dept. of Astronomy, University of Wisconsin, Madison WI 53706 \\
email: ted@noao.edu}
\authoraddr{National Optical Astronomy Observatories, 950 N. Cherry Av., Tucson, AZ 85719}

\author{G. D. Bothun}
\affil{Dept. of Physics, University of Oregon, Eugene OR, 97403 \\
email: nuts@moo2.uoregon.edu}

\and

\author{R. A. Schommer}
\affil{Cerro Tololo InterAmerican Observatory, La Serena, Chile \\
email: rschommer@noao.edu}

\begin{abstract}

We discuss the luminosity function of SNe Ia under the assumption that
recent evidence for dispersion in this standard candle is related to
variations in the white dwarf mass function (WDMF) in the host galaxies.
We develop a simple parameterization of the WDMF as a function of age of a
stellar population and apply this to galaxies of different morphological
types. We show that this simplified model is consistent with the observed
WDMF of Bergeron \etal (1992) for the solar neighborhood.  Our simple
models predict that WDMF variations can produce a range of more than
1.$^m$8 in M$_B$(SN Ia), which is comparable to the observed value using
the data of Phillips (1993) and van den Bergh (1996).  We also predict a
galaxy type dependence of M$_B$(SN Ia) under standard assumptions of the
star formation history in these galaxies and show that M$_B$(SN Ia) can
evolve with redshift.  In principle both evolutionary and galaxy type
corrections should be applied to recover the intrinsic range of M$_B$(SN
Ia) from the observed values.  Our current inadequate knowledge of the
star formation history of galaxies coupled with poor physical
understanding of the SN Ia mechanism makes the reliable estimation of
these corrections both difficult and controversial.  The predictions of
our models combined with  the observed galaxy and redshift correlations
{\it may} have the power to discriminate between the Chandrasekhar and the
sub-Chandrasekhar progenitor scenarios for SNe Ia.  

\end{abstract}


\section{Introduction}

It is generally assumed that type Ia supernovae (SNe Ia) are thermonuclear
explosions of degenerate white dwarfs near the Chandrasekhar limit (cf.
Wheeler \& Harkness 1990), or perhaps mergers of white dwarfs (WDs) in a
binary system (Paczynski 1985). Either detonation or deflagration models
(Arnett 1969; Nomoto \etal 1976) then produce the visible energy release
that characterize the SN light and velocity curves. Detailed models show
considerable differences in these scenarios (Khokhlov \etal 1993), but
their large intrinsic luminosity coupled with the assumed universal
physics involved in the Chandrasekhar mass limit have led to strong
statements concerning the use of the magnitude at maximum (M$_B$(max)) of
SNe Ia as distance indicators (Branch \& Tammann 1992).

There has been some recent evidence that a moderate to large dispersion
exists in M$_B$(max), however, and perhaps in the intrinsic color of these
objects at maximum. This evidence has been revealed by the extensive efforts of
several groups to obtain high quality and frequently sampled observations
of a large number of SNe.  In particular, Phillips (1993) has
summarized high quality SN Ia measurements in nearby galaxies, with the
result that the intrinsic dispersion in M$_ B$(max) appears to be
$\sim 0.^m8$, and correlated with the decay time of the light curve.
Furthermore, the underluminous nature of  SN1991bg (Filippenko
\etal 1992; Leibundgut \etal 1993) is striking.  This apparent dispersion
in M$_B$(max) has led to some discussion of different models of the origin 
of SN Ia explosions.  Unfortunately, current measurements are insufficient to
discriminate in detail between various explosion models (detonation versus
deflagration, etc.) or origin scenarios (see Wheeler \& Harkness 1990;
Iben \& Tutukov 1991; Khokhlov \etal 1993).  Woosley \& Weaver (1994) and
Livne \& Arnett (1995), among others, have discussed explosions of
sub-Chandrasekhar limit WDs.  Pinto \& Eastman (1997) examine the physics
of SN Ia light curves in detail and use an analytic model to study the
sensitivity of the resultant light curves to various properties of
supernova explosions. They find that a variation in total mass can lead to
a sequence of light curves that reproduces the luminosity -- decline rate
relation. Other possible parameters (explosion energy, $^{ 56}$Ni mass,
and opacity) lead to relations between luminosity and light curve shape
that are opposite to the observed behavior. They conclude that the total
mass of the explosion is a natural and simple explanation of the
observations.

If a wider range of masses could be contributing to SN Ia explosions, then
the possibility arises that different stellar populations would produce
different origin functions (see also Kenyon \etal 1993).  Moreover, since
stellar populations do evolve, SN Ia luminosities may depend on the
particular form of the WDMF which should evolve with redshift.  In
principle, these dependencies can produce biases and selection effects in
surveys for distant SNe which become important to evaluate and remove.
These include the standard Malmquist bias concerns, as well as concerns
about galaxy type and position dependencies in the {\it detected} SN
samples.\footnote{For example, a radial gradient in luminosities of SNe in
galaxies could easily exist because of abundance and age gradients, and
combined with the reddening distribution, systematic luminosity
differences for SNe in the outer regions of spiral or irregular galaxies
might dominate the observed samples.}  These concerns are standard ones
with any extragalactic sample, although they have not been extensively
investigated in the SN studies to date, partly due to the small sample
size (see Ruiz-Lapuente \etal 1995).

At issue here is whether the progenitors of SNe Ia have a significant
range in mass, that in turn produces a range in SN Ia luminosities, or if
SNe Ia principally come fron Chandrasekhar mass white dwarfs.  In this
paper, we focus attention on the first possibility and produce a series of
models which produce different white dwarf mass functions (WDMFs) for
differing star formation histories.  These models have predictive power
for both the range in SN Ia luminosities as well as the mean SN Ia
luminosity for a given mean stellar population age.

However, regardless of the physics that produces SNe Ia, it is now well-established
that empirical corrections to their luminosity based on the form of the
light curve (e.g., Riess \etal 1995,1996; Hamuy \etal 1995, 1996b)
produces a Hubble diagram which is linear out to z = 0.1 with a scatter of
$\leq 0^{m}.15$.  To first order, this argues that the intrinsic range of
SN Ia luminosities is irrelevant as the multi-color light curve (MLCS)
and/or luminosity-decline correlations empirically correct for this range.
In fact, these empirical corrections may be so good that systematic
differences between galactic stellar populations may now be revealed.
There is already some observational evidence bearing on this (Hamuy \etal
1996) and so the principle task of our modelling procedure is to
demonstrate how systematic differences in galactic stellar populations
directly lead to systematic galaxy-galaxy differences in SN Ia
luminosities; thus the observed galaxy correlations {\it may} have the
power to discriminate between different progenitor models.

In fact, we will demonstrate that our model predicts evolutionary
corrections to SN Ia luminosities that, at z = 0.5, are an important
percentage of the total cosmological signal differential between q$_o$ = 0
and q$_o$ = 0.5.  In light of the concerted efforts being made in the
detection of SNe Ia at redshifts $\geq$ 0.3 (Perlmutter \etal 1995) and
the expectation that fundamental cosmological parameters can be determined
it seems especially important to understand, in as much detail as
possible, the dependence of mean SN Ia luminosity on the underlying
stellar population.

The thrust of this paper evaluates the WDMF and its variation
as a function of stellar population and evolutionary state, under the
assumption that the dispersion of M$_B$(SN Ia) is correlated with the mass
of the WD progenitor.  In particular we focus on one issue: how does the
WDMF depend on its parent stellar population (Section 2)?  We then apply a
simple parameterization of that dependence to two cases of interest: SNe
Ia arising in different galaxy types (populations with different star
formation histories), and the dependence of progenitor mass on
cosmological look back time (Section 3).  We evaluate these effects on the
determination of $q_o$ from $z \geq 0.3$ SN Ia detections.  Our concern is
in the {\it scatter} in the candle, and the zero point of the flux scale
is irrelevant for this discussion (but relevant for $H_o$).

\section{A Simple Parameterization of the White Dwarf Mass Function}

We consider the luminosity distribution of SNe Ia to be a separable
function, $ \Lambda $, which can be written as

\begin{equation}
   \Lambda = G(m) \ N_{wd} \ L(m_{wd}),
\end{equation}

\noindent
where $G(m)$ is a source function, i.e. a restriction beyond the stellar
population inputs on the mass range of WDs which can become SNe Ia and
which would include various pathways (binary formation, etc.), $N_{wd}$ is
the number distribution given by the WDMF, and $L(m_{wd})$ is the
conversion from WD mass to luminosity (essentially the Ni mass core of the
exploding WD).  For this paper we will assume that $G(m) = 1$ (i.e., no
additional restrictions beyond those which we model); for the
Chandrasekhar mass ignition model, this function would be a delta function
at $1.4 M_{\sun}$.

We build a simple model of the WD mass distribution as a function of
population age based on prescriptions for the WD initial mass -- final
mass relation, theoretical stellar lifetimes, and a star formation rate
(SFR) parameterization.  We assume that the rate of formation of stars of
a given mass at a given time can be characterized by a separable initial
mass function (IMF) and SFR, as

\begin{equation}
   R(m,t) = \Phi(m) \ (A/t_s) \ e^{-t/t_d},
\end{equation}

\noindent
where {\em m} is mass in solar units, {\em A} is a dimensionless
normalization, $t_s$ is the dimensional time unit, $t_d$ is the decay time
of the SFR, and $\Phi(m)$ is the IMF of the form

\begin{equation}
   \Phi(m) = N_o \ m^{\alpha}.
\end{equation}

The number of stars which will leave the main sequence to become WDs in a
given mass interval, $dm$, and in a given time interval, $dt$, is

\begin{equation}
   dN_{evol}(m,t) = \Theta(t - \tau(m)) \ R(t - \tau(m))\, dm\, dt,
\end{equation}

\noindent
where $\tau(m)$ is the timescale of evolution for a star of mass $m$, and
$\Theta$ is a step function equal to $1$ for $t \geq \tau(m)$ and $0$
otherwise, and which allows the two cases of $t < \tau(m)$ and $t \geq
\tau(m)$ to be compactly written.  To determine the number of stars which
have evolved into WDs by a given time, $t$, the above equation is
integrated over $t$ to yield

\begin{equation}		                
   N_{evol}(m,t) = \Theta(t-\tau(m)) \int_{\tau(m)}^t dt \, \Phi(m) \ (A/t_s) \ e^{-(t-\tau(m))/t_d} \ dm.
\end{equation}

\noindent
Letting $ t' = t - \tau(m)$, then

\begin{eqnarray}
   N_{evol}(m,t)& = &\Theta(t-\tau(m)) \ \Phi(m) dm \int_0^{t-\tau(m)} dt' (A/t_s) \ e^{-t'/t_d} \\
                & = &\Theta(t-\tau(m)) \ \Phi(m) dm \ A \ (t_d/t_s) \ [1 - e^{-(t-\tau(m))/t_d}].
\end{eqnarray}

To convert $N_{evol}(m,t)$ to $N_{wd}(m,t)$ requires an initial mass --
final mass relation, which we achieve from a quadratic parameterization of
empirical relation ``A'' of Weidemann \& Koester (1983):

\begin{equation}		                
   M_{wd} = 0.48 - 0.016 \ m + 0.016 \ m^2,
\end{equation}

\noindent
where $m$ is the initial main sequence mass of a star as used above, and
$M_{wd}$ is the mass of the resulting WD.  This agrees with observations,
which are very limited, and gives a $1.376 M_{\sun}$ WD for $m = 8
M_{\sun}$, which we assume to be the highest mass star that produces a WD
remnant.  Clearly, all results we obtain subsequently stem from this
parameterization, and so, to the extent that it can be justified by the
observations, we have a reasonably firm foundation.

We also require $\tau(m)$, the pre-WD evolutionary timescales as a
function of mass, which we achieve from a re-parameterization of the
equations given in Eggleton \etal (1989).  We simplify their
parameterization as we require only lifetimes for stars with masses from
$1$ to $8 M_{\sun}$, whereas their equations are valid for $1$ to $80
M_{\sun}$.  Additionally, we renormalize their stellar lifetimes so that a
$1 M_{\sun}$ star has a main sequence lifetime of $10^{10}$ years.
Following Eggleton \etal (1989) we also take the post main sequence
lifetime to be 15\% of the main sequence lifetime.  The resulting
parameterization is then

\begin{equation}		                
   \tau(m) = t_o \ m^{-2.8},
\end{equation}
   
\noindent
with $t_o = 1.15 \times 10^{10}$ years, for $m = 1$ -- $8 M_{\sun}$.  For
an $8 M_{\sun}$ star, the timescale of evolution is $34$ Myrs.

To use the above equations we set A = $1$ (arbitrary normalization) and
$t_s = 1$ Gyr (i.e., all time units in Gyrs).  We then choose various SFR
models with $t_d = 1, 3, 5, 10,$ and $100$ Gyrs to simulate the range from
single age ellipticals to constant star formation spirals.  We explore the
range $\alpha = 0$ to $-3$ ($\alpha = -2.35$ is the Salpeter value) for
the slope of the IMF and then calculate $N_{evol}$ over the range of time
until t = $12$ Gyrs.  Finally, we transform $N_{evol}$ to $N_{wd}$ via the
initial mass -- final mass relation.

Figure 1 shows the resultant WDMF for different values of the mass
function slope ($\alpha = -3, -2, -1, 0$) in each panel, for two different
SFR decay times ($1$ Gyr, essentially a burst; and $100$ Gyrs, almost
constant SF), and for $6$ different ages ($0.5, 1, 2, 4, 8,$ and $12$
Gyrs) since the onset of SF.

One immediate test of these models is a comparison with the solar
neighborhood WD mass function.  Figure 2 shows the observed mass function
of Bergeron \etal (1992).  As they note, this magnitude-limited survey
selects against the fainter, low radius (high mass) WDs.  Additional
selection may also be caused by the quicker cooling of higher mass WDs,
and possible scale height inflation that would preferentially select
against all stars of higher mass than the current turn-off mass of the
disk population.  Plotted on this distribution is our $10$ Gyr, steady
SFR, $\alpha = -2.35$, model with an  arbitrary normalization. The
agreement is satisfactory after noting that Bergeron \etal (1992)
interpret the lowest mass WDs (first several bins) as likely results of
binary evolution.  On this basis, we believe that our models produce WDMFs
that are astrophysically reasonable.

\section{Luminosity Functions: Predictions, Samples, and Biases}

Woosley \& Weaver (1994) explored the details of $0.6$ -- $0.9 M_{\sun}$
WDs accreting from a companion post main sequence star.  They found a
number of scenarios where $0.1$ to $0.2 M_{\sun}$ of material (He) could
be accreted before a thermal runaway in the surface layers occurred.
Since these thermal runaways propagate more rapidly around the surface of
the WD than the resulting shock wave propagates into the interior of the
WD, the shock wave is focused in the deep interior, often resulting in a
detonation.  We use their models as the basis of our parameterization of
the amount of light given off by the supernova (based on $^{56}$Ni
production) as a function of mass of the accreting white dwarf.  The
Woosley \& Weaver models are meant to explore a range of accretion rates
and metallicities, and we parameterize their results as model A, which has
a mass accretion rate of $2.5 \times 10^{-8} M_{\sun}$ yr$^{-1}$, and
model B, which has a mass accretion rate of $3.5 \times 10^{-8} M_{\sun}$
yr$^{-1}$.  Model A creates SN Ia type explosions for a pre-accretion mass
as low as $0.6 M_{\sun}$, whereas Model B creates SNe Ia for masses as low
as $0.7 M_{\sun}$.  The upper mass limit of their pre-accretion WDs is
$0.9 M_{\sun}$, but we will assume that this relation can be extrapolated
up to $1.1 M_{\sun}$, which is a likely upper limit to C-O WDs (Iben \&
Webbink 1989; but see Kippenhahn \& Weigert 1990).  Our extrapolation and
the unknown upper limit is overly simplistic, but is sufficient for our
purposes.  Our parameterizations of these two models are then

\begin{equation}		                
   L_a(m_{wd}) \sim m^{ni}_a(m_{wd}) = -1.2 + 2.4 \ m_{wd} \ for \ m_{wd} \geq 0.6 \ and
\end{equation}

\begin{equation}		                
   L_b(m_{wd}) \sim m^{ni}_b(m_{wd}) = -1.3 + 2.3 \ m_{wd} \ for \ m_{wd} \geq 0.7,
\end{equation}

\noindent
where $^{56}$Ni masses in excess of $1.376 M_{\sun}$ are set to $1.376
M_{\sun}$.  This adjustment only affects WDs in the incremental mass range
$1.07$ -- $1.10 M_{\sun}$, and only for model A.

The resultant luminosity functions, $\Lambda$, from the product of $N(m)$
and $L(m_{wd})$, are shown in Figure 3 for several combinations of age,
$\alpha$, and SFR parameterizations. The $\Lambda$ functions are very
flat, as expected from the nature of the almost power law mass functions
and the simple linear relation between the Ni mass and the WD progenitor
mass.  These luminosity functions are strongly non-gaussian, which is
likely to be the result in general if the wide mass range assumption we
have made (essentially the Woosley \& Weaver models) are not given any
features by the source function, $G(m)$.

Ideally we would now like to rigorously compare Figure 3 with the observed
SN Ia luminosity function.  However, it is our contention that the
observed LF is poorly known.  Essentially all extant surveys have to be
corrected for completeness.  These incompleteness corrections depend on
the assumed intrinsic form for $\Lambda$\footnote{So $\Lambda_{obs} =
S(\Lambda)$, where $S$ is a selection function which depends on the
magnitude limit and other properties of the survey.}.  As such these
corrections usually have the flavor of self-fulfilling prophecies: the
derived ``$\sigma$'' will depend on the assumed dispersion.  Since the
number of well-measured SNe Ia occurring in host galaxies with well
measured distances is quite low (e.g., the 9 objects in Phillips 1993),
neither the intrinsic LF nor a reliable estimate of the mean M$_B$(SN Ia)
can be made from extant data.

As a result of data paucity, the construction of the proper SN Ia LF is
currently an ambiguous and contentious issue which remains unresolved.
Figure 4a shows the B luminosity function for 29 SNe Ia from the
Cal\'an/Tololo survey (Hamuy \etal 1995, 1996a), plus the 9 objects from
Phillips (1993).  One of our referees argued that the 9 SNe from Phillips
(1993) ``over represents'' peculiar SNe Ia.  We feel this reasoning was
circular since the criteria for inclusion in the Phillips sample is only
that a good distance to the galaxy has been derived (from surface
brightness fluctuations or Tully-Fisher measurements).  How could 
selection based on the existence of an
independent distance estimate cause an over representation of anomalous
SNe Ia?  In fact, the Phillips criteria is exactly what should be used in
the construction of a representative LF as long as no identifiable bias
exists in the distance determinations to these 9 galaxies.   We also
choose the Cal\'an/Tololo sample because it is the largest collection of
SNe Ia with homogeneous (although still not quantified) selection
criteria.  Figure 4a presents the observed LF, uncorrected for any
probable selection effects.  The Phillips (1993) sample of nearby SNe Ia
in galaxies has unknown selection effects, while the Cal\'an/Tololo sample
of southern SNe has some galaxy type dependencies with distance that are
still being explored.  For this sample, absolute magnitudes of SNe Ia are
assigned using redshift as the distance indicator.  The most significant
aspect of Figure 4a is not its shape or mean value but rather the total
luminosity range that is exhibited.  The very faint object evident in this
figure is SN1991bg, a very red SN that has been universally tagged as
being an anomalous SN.

Figure 4b shows the SN Ia LF for all the SNe from Vaughan \etal (1995)
(hereinafter VBMP) with data obtained after 1970 ($30$ SNe).  If we
exclude from the first sample SN1991bg, the two distributions are
essentially identical.  For the 30 objects in Figure 4b, the mean
B-magnitude is $-18.50 \pm 0.49$ while the mean B-magnitude for the $37$
objects in Figure 4a is $-18.50 \pm 0.4$.  These dispersions are
relatively large, and obviously uncorrected SNe Ia would not appear to be
a premier distance indicator.  VBMP claim to be able to lower this
dispersion by identifying and removing SNe with deviant red or blue
color.  Since the intrinsic spectral energy distribution of SNe Ia is not
yet well known from theory, a color-based rejection criteria is at best
risky.  If we examine the 20 objects in the VBMP sample that were
discovered after 1980 and reject SN1991bg and SN1986G (as obvious
deviants), the mean magnitude is $-18.40 \pm 0.46$ (a change of $-0.^m1$
in the mean is significant in the cosmological context).  VBMP reject $3$
more objects from this sample, including the very well studied object
SN1989B.  VBMP specify the observed color (B$-$V = $0.30$) of SN1989B as
being anomalous but Wells \etal (1994) attribute its color to a large
reddening, specifically E(B$-$V) = $0.37$.  Removing this single object
from the 18 most recent SNe in the VBMP sample lowers the dispersion from
$0.46$ to $0.34$ mag!  Yet if reddening is the reason for the anomalous
color, then obviously the absolute magnitude of SN1989B is substantially
brighter than the value listed in VBMP.  After trimming of the anomalous
objects in Figure 4b, VBMP find a distribution with a mean magnitude of
$-18.54 \pm 0.35$.  This mean is very similar to the values we derive for
Figure 4a.

We contend that the SN Ia LF is simply not yet well-determined due to
limited sample sizes and survey volumes, and the difficulty of determining
direct and independent distances to many of the host galaxies.  While we
may have a reasonable estimate for the maximum brightness of SNe Ia, we do
not know the entire LF.  Furthermore, the SN Ia LF of VBMP is not
representative of the {\it whole distribution} of SN Ia luminosities but
rather of VBMP's selected sub-sample in which they have chosen to ignore
or reject a fair percentage of the faintest SNe.  Are these rejected
objects not, therefore, SNe Ia?  Without an adequate explanation of the
mechanism that causes the rejected objects to be anomalously
underluminous, it seems premature to a priori exclude them when specifying
the intrinsic range of SN Ia luminosities and then claim that the sample
of distant SNe Ia is {\it identical} to the selected sub-sample.

For example, we know that the Cal\'an/Tololo sample has obvious selection
effects; SNe Ia fainter than $-18.0$ will not be found in at least half
the surveyed volume, since they fall below the apparent magnitude cutoff
of the survey.  Some of these selection effects are discussed in Hamuy
\etal (1994).  While we are not prepared here to analyze completeness of
extant SNe samples, we schematically illustrate
our concerns in Figure 5, which shows the redshift distribution
of the Cal\'an/Tololo SN Ia sample. This distribution is extremely flat,
with a median recessional velocity of $\sim$ 14,000 km s$^{-1}$, but
extending out past 30,000 km s$^{-1}$.  We have included lines in this
figure to demonstrate the expected increase in the sample due to volume
effects. The dashed line is a normalization assuming the survey is
complete out to 4,000 km s$^{-1}$ (which {\it may} be representative of
the lower luminosity SNe Ia), while the dotted line assumes it is complete
out to 14,000 km s$^{-1}$ (representative of the brighter SNe Ia).  In
either case, we conclude that the sample is severely incomplete through
much of its volume; in the first case it is 98\% {\it incomplete} at the
median redshift of 14,000 km s$^{-1}$. If the deeper completeness
normalization is assumed, the excess of low redshift SNe are those of
fainter absolute magnitude. These comments about incompleteness in the
samples are a reflection of our concerns about the completeness and
accuracy of the extant SN samples which have been used to construct the SN
Ia LF.

Our simple model of the {\it range} of SN Ia luminosities attempts to
explore the systematic connection between this range and galaxy type.
Indeed, it may be very difficult to judge if the very distant SNe Ia have
a broad or narrow distribution of absolute magnitudes, due to selection
effects and cosmological corrections.  We are exploring a scenario that
makes certain predictions that can be tested on both local and distant
samples.

\subsection{Galaxy Population Dependencies}

For the purposes of this discussion, we consider that our simple models
may be assigned to galaxy types based on stellar population type.  Thus we
will assume that we can characterize elliptical galaxies as single burst
models, with the majority of stars formed $12$ Gyrs in the past, and with
a $1$ Gyr exponential decay time.  We will also assume a IMF slope of
$\alpha = -2$ for this single burst model.  We will assume an actively
star forming galaxy (SFG) can be characterized by star formation starting
approximately $8$ Gyrs ago (the approximate age of the Galactic disk) with
effectively continuous star formation ($t_d = 100$ Gyrs) and with several
IMF slopes. Clearly, these assumptions can be challenged, but they
correctly predict the average UBV color differences between spiral and
elliptical galaxies (see Larson \& Tinsley 1978; Bothun 1982).  For the
cosmological parameters we will assume $H_o = 50$ km s$^{-1}$ Mpc$^{-1}$
(after all this is a paper on SNe) and $q_o$ = 0.5.  This universe is less
than $14$ Gyrs old.

Although the power law nature of the LFs shown in Figure 3 precludes the
calculation of a reliable mean SN Ia luminosity as a function of galaxy
type, we can use these means to make rough estimates.  Figure 6 shows the
mean SN Ia luminosities that are obtained by integrating over the Ni mass
distribution as normalized by total number of SN Ia events.  This figure
demonstrates that our model LFs do not change shape after $\sim 0.5$ Gyrs,
which is the evolutionary timescale of the lowest mass progenitors ($3
M_{\sun}$) which explode as SNe Ia.  We caution, however, that our model
does not include the unknown, but possibly large, time delays inherent in
the binary mechanism before mass transfer begins.  Thus the large change in
the SN Ia LF shape evident at early times should take place over a greater
time period, making the SN Ia LF more sensitive to stellar population age
than this figure implies.  Figure 6 demonstrates that, in the case in
which ellipticals and spirals have the same IMF slope $\alpha$, the
differences in mean SN Ia luminosity are small ($\leq 0.04$ mag).  This
small difference is not surprising as star formation histories with
similar $\alpha$ will produce very similar WDMFs once the mean age is
greater than the evolutionary timescale of the WD progenitors.  The
difference in mean SN Ia magnitude between an $\alpha = -2$ elliptical and
an $\alpha = 0$ spiral is $0^m.22$.  (An even more extreme difference
occurs with large lookback times.)  The data from the SN samples support
these dependencies on the underlying stellar population.  Hamuy \etal
(1996a) find that the brightest SNe occur in late type galaxies (see their
Figure 3) and even more strikingly, that a strong correlation exists
between the decline rate of the SN Ia light curves and the host galaxy
morphology (see their Figure 4).  Branch \etal (1996) find that SNe Ia
that occur in ``red'' galaxies are 0.3 magnitudes less luminous than those
that occur in ``blue'' galaxies.   While our model calculations are meant
to be illustrative only, they do show that differences in mean stellar
population age and/or slope of the IMF can produce significant differences
in mean M$_B$(SN Ia) that are approximately the same size as the effects
seen in existing data samples.

However, variations in mean M$_B$(SN Ia) between galaxy types are not the
relevant quantity with respect to distance measurements.  Rather, the
total range of M$_B$(SN Ia) is important, especially when considering the
effects of Malmquist bias.  In Figure 7 we plot the initial mass -- final
mass relation (equation 8) together with nickel mass production.  The
models indicate a range of $4$ -- $5.5$ in Ni mass, which indicates a
range in SN Ia luminosities of up to $1^m.8$.  We terminate our masses at
a $1.4 M_{\sun}$ WD, but if WD mergers at all masses are a possible
channel, then the functions should be continued up to the possible sum of
$2.8 M_{\sun}$, giving a total possible range of $2.6$ mags for SN Ia
luminosities.

Our predicted range is similar to the observed range shown by Phillips
(1993).  However, this comparison is only indirect.  Our predicted results
are for the SN Ia luminosity range in a single galaxy for a specific WDMF,
whereas in comparing to observations, we are sampling over a range of
galaxy types.  Still, the rough agreement between the results based on our
model parameterization and the available observations has an alarming
implication:  an order of magnitude range in M$_B$(SN Ia) immediately
suggests that Malmquist corrections are large for any distant
extragalactic sample of SNe.  We believe the current observations are
effectively sampling this range.  That the method of Riess \etal (1996)
can lead to Hubble diagrams with such low dispersions indicates that the
light-curve corrections to SN Ia luminosities are very effective at
compressing this intrinsic luminosity range.  If these corrections
continue to work well in larger samples, then it becomes clear that the
intrinsic luminosity range of SNe Ia is essentially irrelevant with
respect to determining distances.  All that is required is a secure
calibration of these light-curve corrected luminosities.

For the simple case in which the Ni mass is proportional to the WD mass
(e.g., equations 10 and 11), our models predict a spread in SN Ia
luminosities of at least $1^m.5$.  However, invoking variations in WD
progenitor mass as the sole cause of the SN Ia luminosity spread may not
be necessary.  Various explosion scenarios can easily give 50\% variation
in the energy release (Khokhlov \etal 1993; H\"oflich \etal 1996).
Whether these models would have a systematic dependency on the WDMF or on
the evolutionary state or metal abundance undoubtedly depends in detail on
the nature of the explosion.  We have also ignored any effects of changes
in the WDMF on the binary frequency or whatever progenitors are the SN Ia
source $G(m)$ (cf.\ Kenyon \etal 1993).  While the wide binary source
function may be independent to zeroth order of the details of the
individual stellar mass function, the SN Ia source function probably
evolved in a more complicated manner than a simple dependence on mean WD
mass.  Hence, several physical effects can cause the SN Ia LF to depart
significantly from a delta function.

\subsection{Cosmology and Evolutionary Corrections}

The redshift -- magnitude relation in standard form yields the equation

\begin{equation}
   m = M + 25 - 5 \ log \ H_o + 5 \ log \ cz + 1.086 \ (1-q_o) \ z + ....
\end{equation}

\noindent
At z$ = 0.5$, for $H_o = 50$ km s$^{-1}$ Mpc$^{-1}$, the difference
between $q_o$ of $0.0$ and $0.5$ (empty versus critical mass models) is
$0.^m27$ (assuming a zero cosmological constant).  With photometric
accuracies of $\sim 0.^{m}1$ per SN event, statistics of a  sample of $10$
well-measured objects would permit an $\sim 8 \sigma$ discrimination
between  empty and critical models.  Clearly, however, systematic errors
or biases at the $10$\% level become very significant and lead to an
effective $q_o$ measurement.\footnote{We would measure $q_o^{eff} = q_o -
(dL/dt)/ L / H_o$, in the case of luminosity evolution, for example.}

For these cosmological parameters the look-back time at $z = 0.5$ is
$3.75$ Gyrs. Inspecting our models we see little change in our E galaxy
sources.  However, the actively SFG shows significant evolutionary effects
in the sense that the WDMF is populated toward the more massive objects in
the past, and thus the $\Lambda$ function produces brighter SNe.  The WDMF
also depends sensitively on the assumed age parameter for the SFG; if the
galaxies are assumed to initiate star formation $8$ Gyrs ago, at $z = 0.5$
the mean luminosity can be as large as $0^{m}.31$ brighter for the $\alpha
= 0$ case.\footnote{We ignore here the K-correction issue, which can be
complicated at the few percent level for complex spectral types such as
SNe.  For example, different K-corrections are probably necessary at
different epochs for a given supernova event, because of the significant
evolution of the spectral energy distribution.  Clearly, shifting the
observed bandpass with redshift is an important consideration (cf.\ Hamuy
\etal 1993; Kim \etal 1996).}

The nature of the general galaxy population at $z = 0.5$ is still poorly
determined.  At a minimum, however, we expect the fraction of young or
starbursting galaxies in some random field to be significantly higher than
is observed at the present epoch.  The SN Ia rate at some epoch, z, is a
function of the total number of young stars that exist at that epoch since
massive white dwarfs are produced by short-lived stars.  Hence, a single,
massive starbursting galaxy could completely dominate the rate.
Additionally, early star formation may be characterized by a far different
IMF slope than we observe today.  This would be particularly troublesome
as the SN Ia LF changes dramatically with IMF slope (see Figure 6).  For
our purposes, we pose one specific question:  how much of a star formation
burst is required to create a significant number of SNe Ia from the burst
population relative to a single-age $5$ -- $8$ Gyr (elliptical) galaxy?
SNe that occur in ellipticals (or the old population in a spiral which we
assume is negligible at these redshifts) can be thought of as being the
background SN population against which SNe occurring in star bursting
galaxies are detected.  The relative contribution of the SF and background
populations can be parameterized as

\begin{equation}
   {No. \ burst \ SNe \over No. \ background \ SNe} = {n_{burst} \ exp (-t/t_o)
   \over n_{back} \ exp (-t/t_o)} = {n_{burst } \ exp (-1 \ or \ -0.5)
   \over n_{back} \ exp (-8 \ or \ -5)}
\end{equation}

\noindent
For population age, t = $5$ Gyrs,

\begin{equation}
   (SN_{burst}/SN_{back}) = (n_{burst}/n_{back}) \ (55 - 90)
\end{equation}

\noindent
and for population age, t = $8$ Gyrs,

\begin{equation}
   (SN_{burst}/SN_{back}) = (n_{burst}/n_{back}) \ (1100 - 1800).
\end{equation}

Table 1 provides the percentage of SNe resulting from a burst as a
function of the percentage of mass in the burst population relative to
underlying stellar population.  Again, the values in this table come from
considering one single age elliptical with one starbursting spiral.
Column 1 lists the percentage of total SNe that come from the starbursting
spiral, while columns 2 and 3 list the starbursting mass fraction for the
t = $5$ and $8$ Gyrs cases.

For the case of low burst strength (e.g., $\leq 2$\%), we find the
expected result that since the field contains two galaxies, $50$\% of the
SNe come from one of the two galaxies.  However, for a star formation
burst of $10$ -- $20$\%, $90$\% of the SNe will come from that one
starburst galaxy.  The situation is even more extreme if we consider a
true starburst galaxy (burst strength $\geq 100$\%) in which case $99$\%
of the SNe come from that one galaxy.  These results indicate that if a
field at $z = 0.5$ contained $90$\% ellipticals and $10$\% starburst
spirals with burst amplitudes of $10$ -- $20$\%, then $50$\% of the total
SNe generated by these galaxies would come from the minority population.
If, however, these galaxies are preferentially dusty, then extinction
effects may reduce the detection of SNe Ia from these hosts.  Thus,
samples of distant SNe might be dominated by host galaxies which have
young mean ages.

\section{Conclusions and Caveats}

Overall, our simple model of the dependence of the SN Ia luminosity on the
underlying WDMF allows us to make the following predictions:

1.  In the mean, the SNe Ia occurring in spiral galaxies should be more
luminous than those occurring in elliptical galaxies; bright SNe Ia in E
galaxies should be very rare.  This effect can be seen in the data
compilations of Phillips (1993) and in the Cal\'an/Tololo survey (Hamuy
\etal 1995, 1996a).

2.  A correlation should exist between SN Ia luminosity and the color of
the host galaxy population, with brighter SNe present in bluer galaxies.

3. The SNe in the disks of spiral galaxies should be more luminous in the
mean than those in the bulges.  This effect may be hard to observe because
of reddening effects.  A reddening independent light curve parameter (such
as $\Delta$m$_{15}$) should correlate with position in a spiral galaxy,
with the broader light curves (smaller $\Delta$m$_{15}$ values)
preferentially in the disks or outer regions of the spirals.

4.  More distant SNe should show slower light curve decay (smaller
$\Delta$m$_{15}$ values) than the nearby sample because these SNe are
preferentially more luminous.  This prediction is a consequence of both
starbursting galaxies dominating distant samples and the Malmquist bias
that directly results from the large range in intrinsic SN Ia
luminosities.

5.  Distant SNe are expected to come predominantly from bright, blue,
spiral or irregular galaxy hosts, most of which are in an elevated state
of star formation.  The mean age of these hosts will be younger than
the mean age of most z = 0 calibrating galaxies, making it important that
starburst galaxies like NGC 5253 are included in the local calibrating 
sample.  We have already shown that $M_B$(max) is sensitive to the mean age of
the stellar population.  Thus correcting for this mean age effect requires 
detailed knowledge of the nature of the stellar populations in distant 
galaxies.  The predicted difference in $M_B$(max) obtained under modest 
assumptions about the star formation history of galaxies is an appreciable 
fraction of the cosmological signal that
distinguishes $q_o = 0$ from $q_o = 0.5$.

6. The form of the LF for SNe Ia should be approximately a power law (see
Figure 3), if we assume the binary formation function introduces no strong
features.  Larger samples of low redshift SNe will be needed to determine
this function.

7. In general, the form of the WDMF predicts a range of 1.5--2.5$^m$ in SN
Ia luminosities.  We have argued that current SN Ia samples have
effectively sampled this range in luminosity and that their usefulness as
a distance indicator depends critically on the universality of the light
curve correction algorithms (e.g., MLCS) in compressing this luminosity
range.

If many of these predictions are borne out, we would contend that such
observational evidence favors the sub-Chandrasekhar mass hypothesis as the
main SN Ia progenitor.  In fairness, our results and modelling procedure
and its application to the SN Ia distance scale are subject to a number of
caveats and we close this paper by discussing them.

In converting WD masses into SN Ia luminosities we use the recent
calculations for sub-Chandrasekhar explosions by Woosley \& Weaver
(1994).  These models accurately reproduce the observed correlation
between decline rates of the light curves and luminosity, and are able to
produce more $^{44}$Ti and $^{48}$Cr than other types of models (see
discussion in Livne \& Arnett 1995), which is important in matching solar
abundances.  The 1D treatment of Woosley \& Weaver (1994) yields similar
results to the 2D treatment of Livne \& Arnett (1995).  We choose the
Woosley \& Weaver models because they have clear predictive power, not
because we consider these models to be definitive.  While models of
Chandrasekhar mass explosions can also yield a range of luminosity (e.g.,
H\"oflich \etal 1995; 1996), based on the nature and degree of turbulence
in the explosion, we contend that, because the dispersion in SN Ia
luminosities is not small and seems to be correlated with galaxy
morphology, effects in addition to explosion physics most likely produce
the observed LF.  It is not our intention to delve into SN explosion
physics or discuss which SN models in the literature are more nearly
correct.  Instead we have argued that a major part of the observed
luminosity range for SNe Ia {\it can} result from a dependence of mean SN
Ia luminosity upon the mean stellar population of the host galaxy.

In this case, the mixture of host galaxy types in any SN Ia sample
determines the LF for that sample.  Thus, it is not surprising that there
is disagreement over the form of a typical host galaxy in
the SN Ia sample at $z = 0$.  Moreover,
our models clearly show the importance of starbursting galaxies in distant
samples.  The higher SN Ia rate in these galaxies allows the minority
population to dominate the observed frequency.  Since these galaxies have
younger mean ages and hence more extended WDMFs, the range of SN Ia
luminosities is larger than that in a $z = 0$ spiral or elliptical.
Indeed, the distribution of SN Ia host galaxies in the nearby Universe
shows some curious properties which makes it hard to determine if the
typical host is a spiral or an elliptical.  For instance, the modestly
star forming galaxy M 100 has had four detected SNe since 1901 (one of
which is a type II), whereas the megastar elliptical M 87 has had zero.
NGC 5253, a low mass but actively star forming galaxy, has had two
detected SNe Ia in the last 100 years.  By comparison, the Coma cluster,
home to $\sim$ $10,00$ gas poor L* galaxies (e.g., $10^4$ NGC 5253
masses), has not had a single SN Ia event detected for the last 22 years.

In contrast to this anecdotal evidence, which suggests that spiral hosts
dominate over elliptical hosts, the 30 or so SNe Ia that have been
detected in the Cal\'an/Tololo survey show nearly equal numbers of
elliptical and spiral hosts beyond $z = 0.033$, demonstrating an {\it
anti-}Malmquist bias. The dominance of nearby spiral hosts [at redshifts $
z \leq 0.033$ (Hamuy \etal 1996a)], may be a result of the avoidance of
nearby clusters in the search fields.  However, in the distant half of the
sample no {\it a priori} selection against clusters existed, and hence,
proportionately more ellipticals should be in that sample, causing some of
the variation.  Thus variation in the S/E host ratio could reflect these
selection criteria, as well as the low space density of relatively
unreddened starburst spirals
in the local universe ($z \leq 0.1$).  It is unlikely, however, that
similar circumstances would continue to hold at larger redshifts.

Finally, we comment on the use of light curve corrections to SN Ia
luminosities in the context of our model.  Astrophysical measurements,
based on either light curve parameters (Phillips 1993; Riess \etal
1995,1996; Hamuy \etal 1996a) or spectroscopic analysis (Nugent \etal
1995), appear to correlate well with peak SN luminosity.  Hamuy \etal
(1996a) show that high quality light curves exhibit a characteristic shape
and luminosity -- decay time relation (Phillips 1993; hereafter the
Phillips relation) that produce significantly improved peak magnitudes and
much more accurate relative distances than the use of a single absolute
magnitude calibration (see also Riess \etal 1995,1996).  In the model
explored here, the Phillips relation represents a stellar mass sequence.
Although we do not attempt to derive the relation between light curve
parameters and mass explicitly, the luminosity -- mass relation is itself
linear.

Intrinsic dispersion around the Phillips relation or the MLCS relation of
Riess \etal (1996) would be caused by additional parameters (e.g.,
metallicity), which may or may not correlate with the host galaxy stellar
population and the WDMF.   It is the dispersion around these relations
that become directly relevant to correcting distant samples for Malmquist
bias.  Moreover, corrections to SN Ia peak luminosities using $z = 0$
light curves may not be strictly applicable to distant SNe in
star-bursting galaxies as such galaxies will be rare (or perhaps
non-existent) in the nearby calibrating sample.  At the very least, our
models show that knowledge of the WDMF as inferred from the nature of the
stellar population of the host galaxy is critical in order to determine
potential systematic differences in light-curve shape between the distant
host galaxy and the calibrating sample.  Minimizing these differences may
well validate the approaches of Hamuy \etal (1995, 1996b) and Riess \etal
(1995,1996) that have produced linear Hubble relations  out to $z = 0.1$
with a scatter of approximately 0$^{m}.13$ -- $0^{m}.17$ (Hamuy \etal
1996b, equations $7$ -- $9$) which raise the expectation that $q_o$ can be
determined from such data.

\acknowledgments

We acknowledge helpful discussions with Karl Fisher, Philip Pinto, and
Michael Richmond.  We also acknowledge Mark Phillips, Mario Hamuy, and
Nick Suntzeff for inspiring us to investigate the possible connection
between SN Ia luminosities and the underlying stellar population.
Finally, we wish to dedicate this paper to Marc Aaronson, who would have
wanted a thorough investigation of the reliability of SNe Ia as standard
candles.

%\eject

\newpage

\figcaption[vonhippel.fig1ab.ps]
{WDMFs for different IMF slopes (panels display exponential SF duration
and IMF slope) and for 6 different ages ($0.5, 1, 2, 4, 8,$ and $12$ Gyrs,
bottom to top, respectively).  a) WDMF of burst models with $1$ Gyr
exponentially decaying star formation; b) WDMF of active star forming
galaxies with $100$ Gyr exponential star formation.}

\figcaption[vonhippel.fig2.ps]
{Observed Galactic WDMF (from Bergeron \etal 1992) plotted versus $10$
Gyr, $\alpha = -2.35$ model for an active star forming galaxy ($100$ Gyr
exponential star formation).}

\figcaption[vonhippel.fig3.ps]
{The differential SN Ia luminosity functions of the cumulative (time
integrated) stellar population as derived in the text, for the burst
models (a, b) and active SFGs (c, d) for $2$ different mass function
slopes ($\alpha = -2, 0$) and 5 different ages ($1, 2, 4, 8,$ and $12$
Gyrs, top to bottom, respectively).}

\figcaption[vonhippel.fig4a.ps,vonhippel.fig4b.ps]
{a) Histogram of absolute B magnitudes for SNe Ia in the sample of
Phillips (1993) and the Cal\'an/Tololo survey (Hamuy \etal 1996a). The
magnitudes of the 9 events from the Phillips paper have been adjusted to a
zero point consistent with the Cal\'an/Tololo sample calibration.  b)
Histogram of absolute B magnitudes for the SNe Ia after 1970 from the
sample from Vaughan \etal (1995).}

\figcaption[vonhippel.fig5.ps]
{The velocity histogram of the Cal\'an/Tololo SN Ia sample.  Shown for
comparison are lines showing the volume increase, normalized to the
observed SN number at cz = 4,000 km s$^{-1}$ (dashed) or cz = 14,000 km
s$^{-1}$ (dotted). Under either assumption, the sample is seriously
incomplete (see text).}

\figcaption[vonhippel.fig6.ps]
{The mean SN Ia luminosity as a function of age, SF history, and IMF slope
for our model A.  The luminosity is in terms of equivalent Ni mass.  The
clustered lines at a particular IMF slope are for models with different SF
histories, with the solid line being the $t_d = 1$ Gyr (burst) model, the
dashed line being the $t_d = 10$ Gyr model, and the dotted line being the
$t_d = 100$ Gyr (steady SF) model.}

\figcaption[vonhippel.fig7.ps]
{The relevant mass ranges for creating model SNe Ia.  The line marked
``WD'' shows the initial mass -- final mass relation, whereas the other
two lines display the Ni mass produced in the SN Ia explosions for models
A and B, as a function of the main sequence mass of the progenitor.}

\clearpage

\begin{table}
\centering
\caption{Burst statistics}
\begin{tabular}{ccc}
     &         &              \\
N from burst & n$_{burst}$/n$_{back}$ at t=5 & n$_{burst}$/n$_{back}$ at t=8 \\
     &         &              \\
50\% & 1-  2\% & $\leq$ 0.1\% \\
90   & 10- 16  & $\leq$ 1     \\
99   & 110-180 & 6 - 9        \\

\end{tabular}
\end{table}


\begin{references}

\reference{} Arnett, D. 1969, \apss, 5, 180

\reference{} Bergeron, J., Saffer, R., \& Liebert, J. 1992, \apj, 394, 228

\reference{} Bothun, G.D. 1982, \apjs, 50, 39

\reference{} Branch, D., Romanishin, W., \& Baron, E. 1996, \apj, 465, 73

\reference{} Branch, D., \& Tammann, G. 1992, \araa, 30, 359

\reference{} Eggleton, P.P., Fitchett, M.J., \& Tout, C.A. 1989, \apj,
347, 998

\reference{} Filippenko, A.V., \etal 1992, \aj, 104, 1543

\reference{} Goobar, A., \& Perlmutter, S. 1995, \apj, 450, 14

\reference{} Hamuy, M., \etal 1994, \aj, 108, 2226

\reference{} Hamuy, M., Phillips, M.M., Maza, J., Suntzeff, N.B.,
Schommer, R.A., \& Aviles, R. 1995, \aj, 109, 1

\reference{} Hamuy, M., Phillips, M.M.,  Schommer, R.A., Suntzeff, N.B.,
Maza, J., \& Aviles, R. 1996, \aj, 112, 2391 (1996a)

\reference{} Hamuy, M., Phillips, M.M.,  Suntzeff, N.B., Schommer, R.A.,
Maza, J., \& Aviles, R. 1996, \aj, 112, 2398 (1996b)

\reference{} Hamuy, M., Phillips, M.M., Wells, L.A., \& Maza, J. 1993,
\pasp, 105, 787

\reference{} H\"oflich, P., Khokhlov, A.M., \& Wheeler, J.C. 1995, \apj,
444, 831

\reference{} H\"oflich, P., Khokhlov, A.M., Wheeler, J.C., Phillips, M.M.,
Suntzeff, N.B., \& Hamuy, M. 1996, \apjl, 472, L81

\reference{} Iben, I., \& Tutukov, A. 1991, \aj, 370, 615

\reference{} Iben, I. \& Webbink, R.F. 1989, in White Dwarfs, ed. G.
Wegner, (Springer-Verlag: Berlin), 477

\reference{} Kenyon, S.J., Livio, M., Mikolajewska, J., \& Tout, C.A.
1993, \apjl, 407, L81

\reference{} Khokhlov, A., Muller, E., \& H\"oflich, P. 1993, \aap, 270,
223

\reference{} Kim, A., Goobar, A., \& Perlmutter, S. 1996, \pasp, 108, 190

\reference{} Kippenhahn, R., \& Weigert, A. 1990, in Stellar Structure and
Evolution, (Springer-Verlag: Berlin), 340

\reference{} Larson, R.B., \& Tinsley, B.M. 1978, \apj, 219, 46

\reference{} Leibundgut, B., \etal 1993, \aj, 105, 301

\reference{} Livne, E., \& Arnett, D. 1995, \apj, 452, 62

\reference{} Nomoto, K., Sugimoto, D., \& Neo, S. 1976, \apss, 39, L37

\reference{} Nugent, P., Phillips, M., Baron,E., Branch, D., \&
Hauschildt, P. 1995, \apjl, 455, L147

\reference{} Paczynski, B. 1985, in Cataclysmic Variables and Low Mass
X-Ray Binaries, eds. D. Q. Lamb and J. Patterson, (Reidel: Dordrecht), 1

\reference{} Perlmutter, S., \etal 1997, \apj, accepted

\reference{} Perlmutter, S., \etal 1995, \apjl, 440, L41

\reference{} Phillips, M.M. 1993, \apjl, 413, L105

\reference{} Pinto, P.A., \& Eastman, R.G. 1997, \apj, submitted

\reference{} Riess, A.G., Press, W.H., \& Kirshner, R.P. 1995, \apjl, 438,
L17

\reference{} Riess, A.G., Press, W.H., \& Kirshner, R.P. 1996, \apj, 473,
88

\reference{} Ruiz-Lapuente, P., Burkert, A., \& Canal, R. 1995, \apjl,
447, L69

\reference{} Tammann, G.A., \& Sandage, A. 1995, \apj, 452, 16

\reference{} van den Bergh, S. 1996, \apj, 472, 431

\reference{} Vaughan, T.E., Branch, D., Miller, D.L., \& Perlmutter, S.
1995, \apj, 439, 558 (VBMP)

\reference{} Weidemann, V., \& Koester, D. 1983, \aap, 121, 77

\reference{} Wells, L.A., \etal 1994, \aj, 108, 2233

\reference{} Wheeler, J.C., \& Harkness, R.P. 1990, Rep.Prog.Phys., 53,
1467

\reference{} Woosley, S.E., \& Weaver, T.A. 1994, \apj, 423, 371

\end{references}
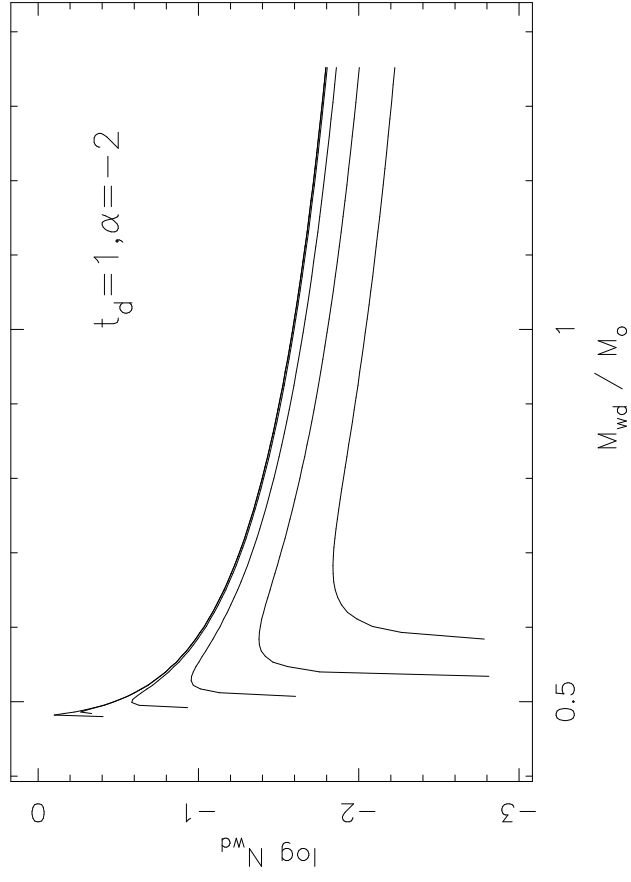
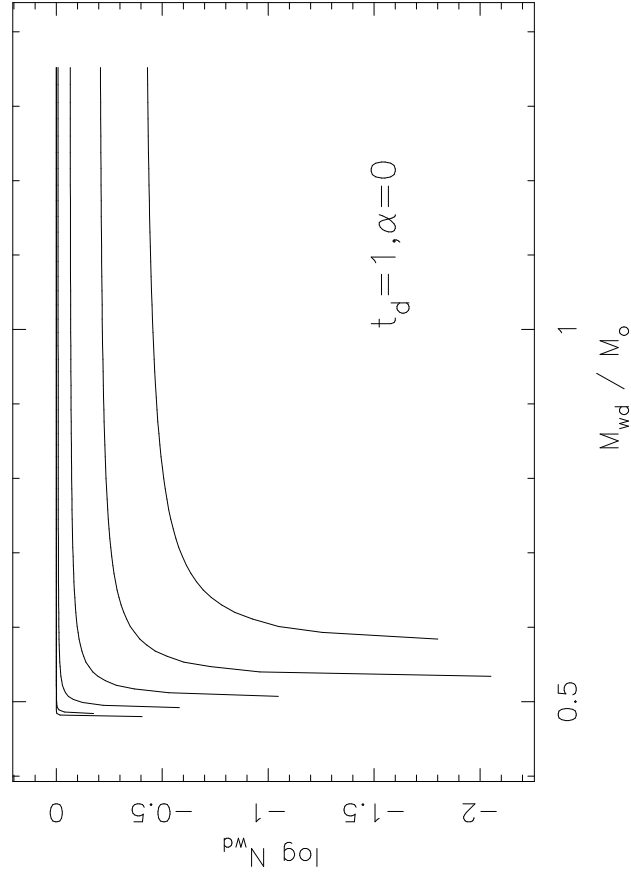
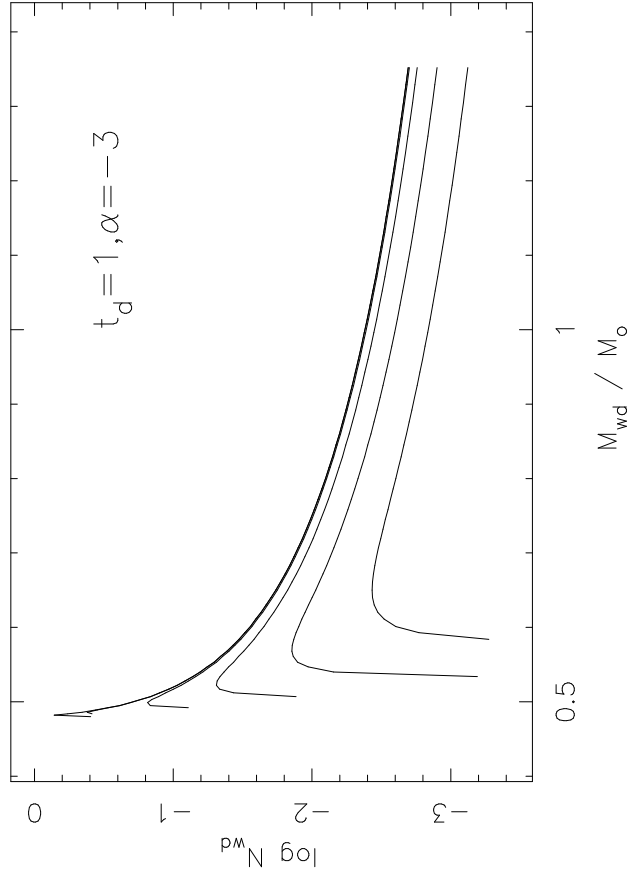
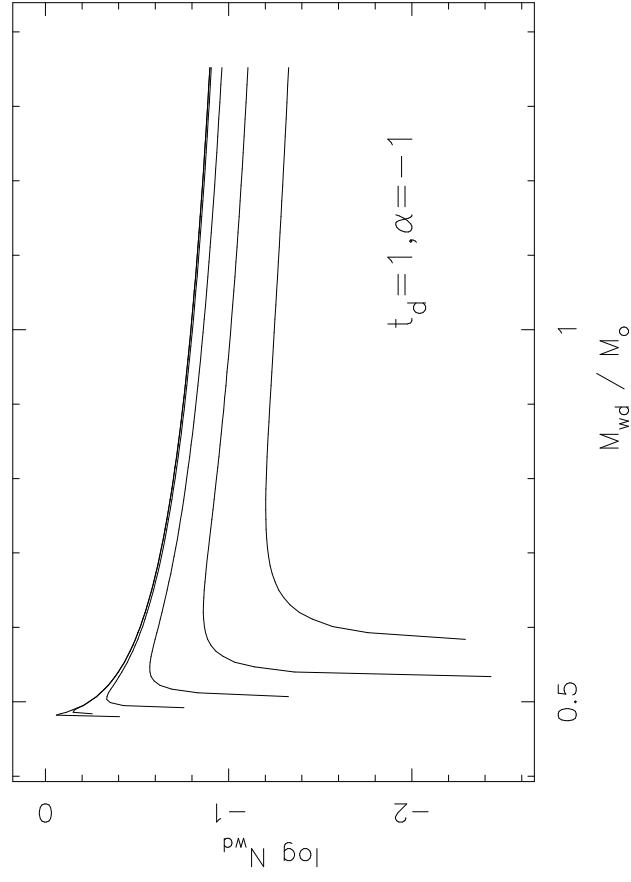
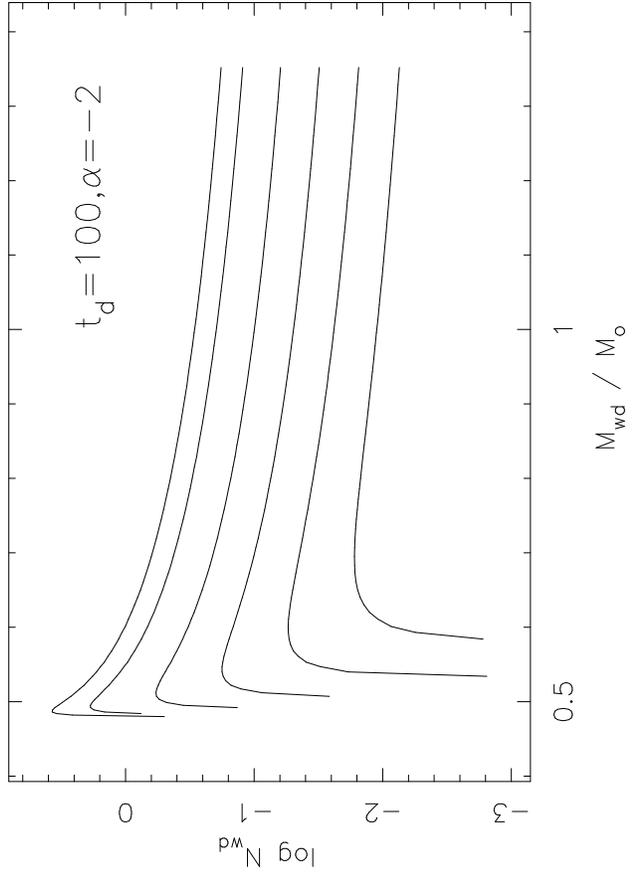
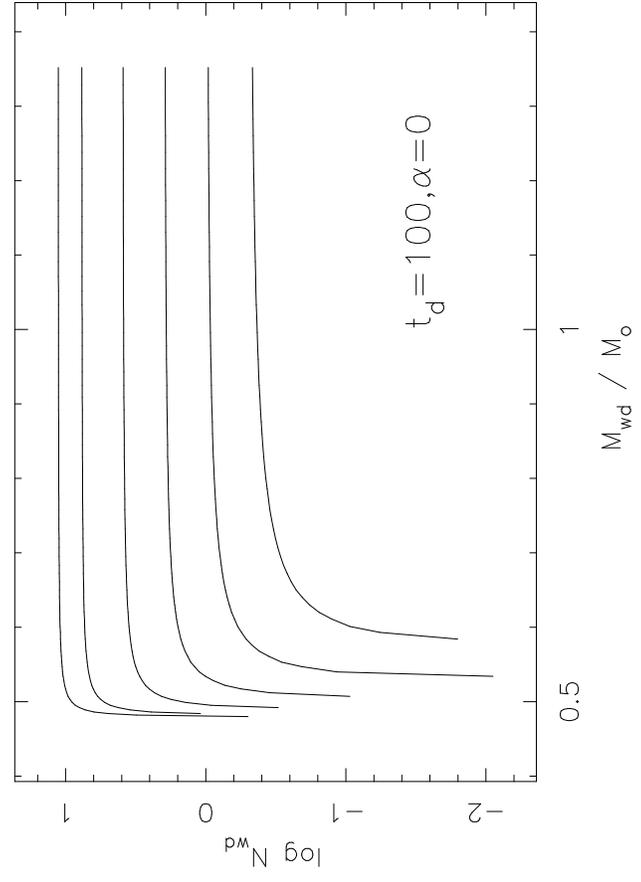
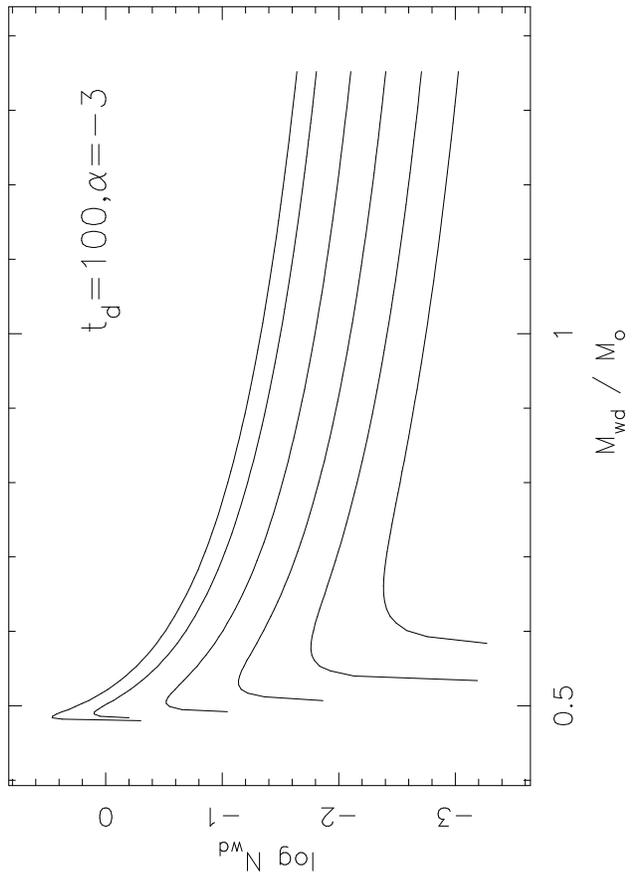
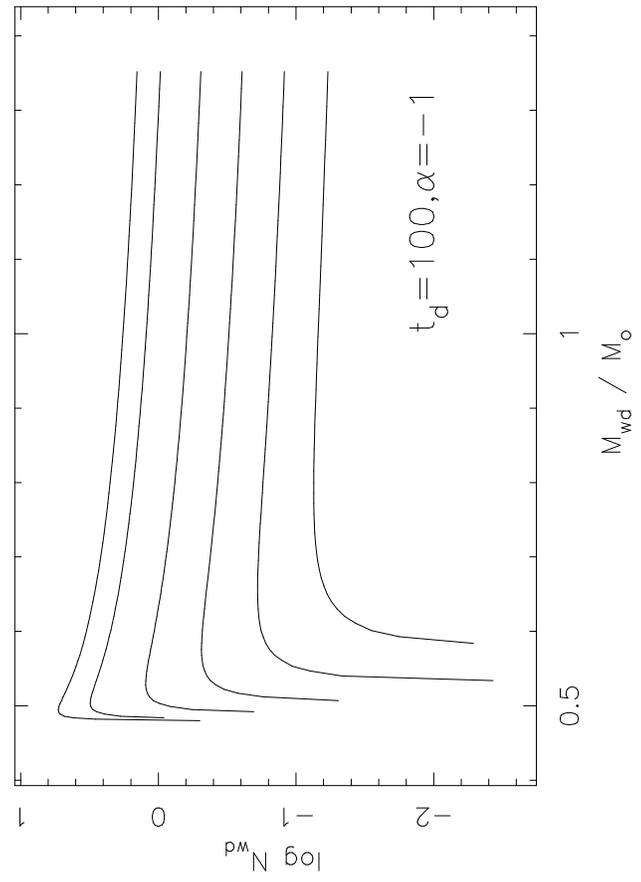
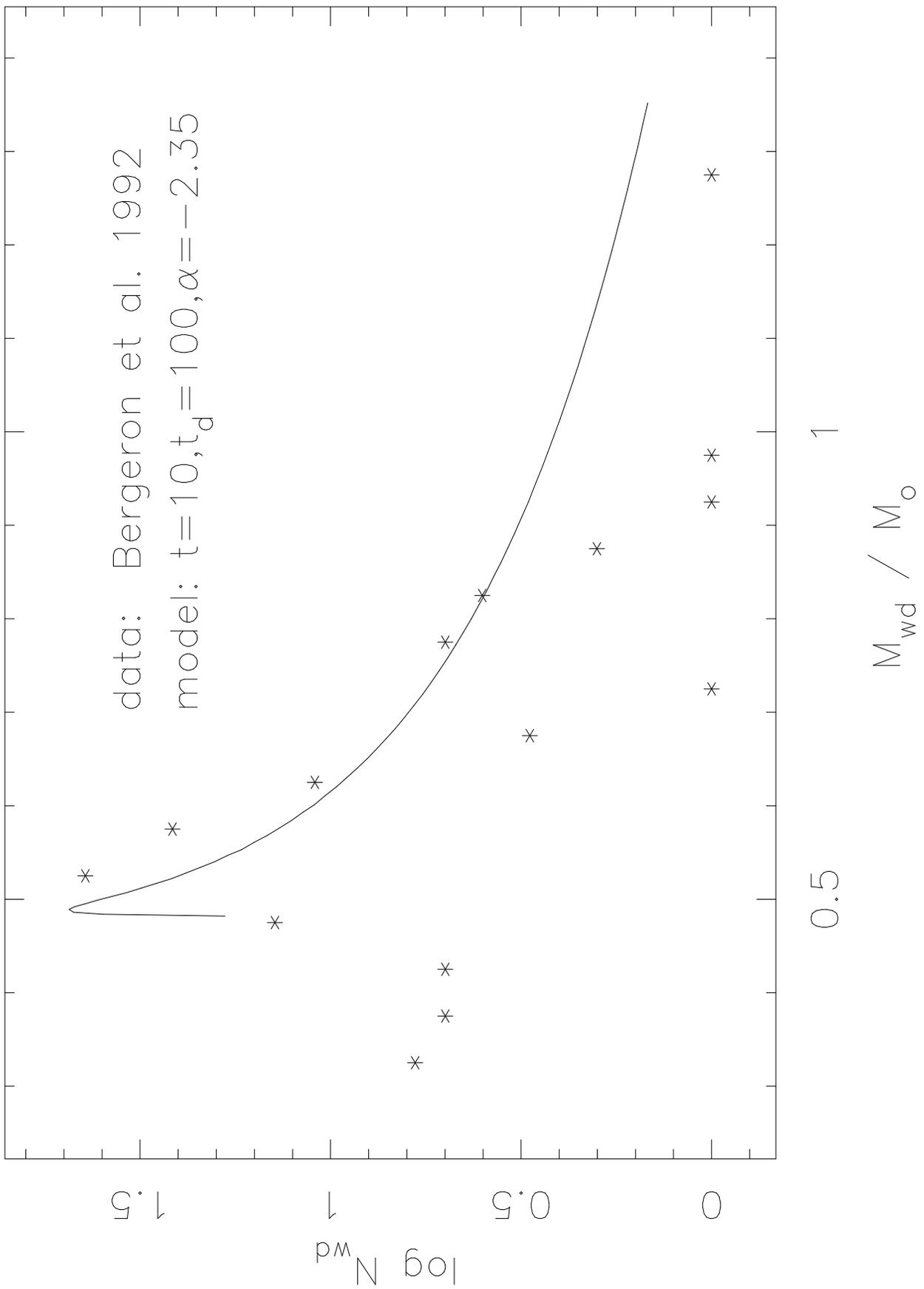
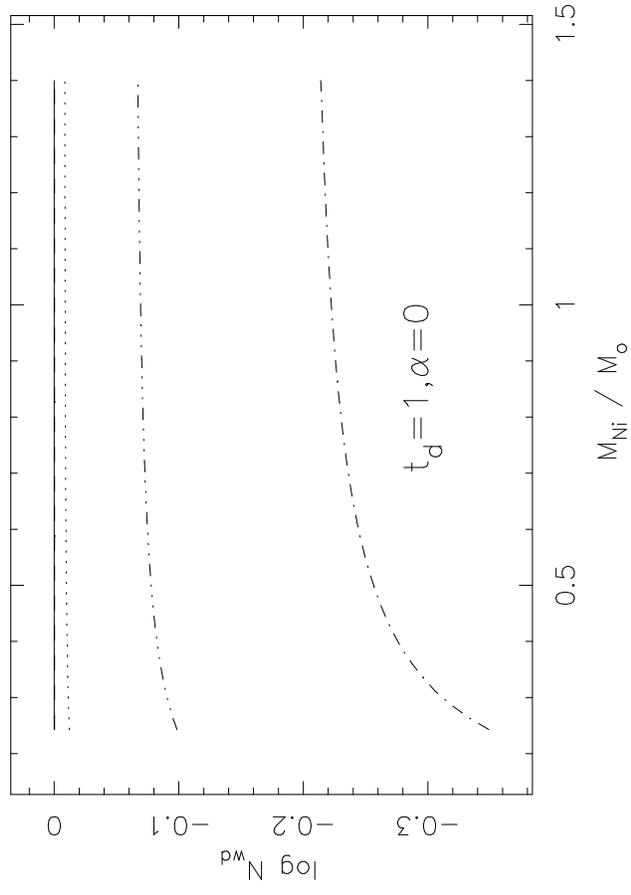
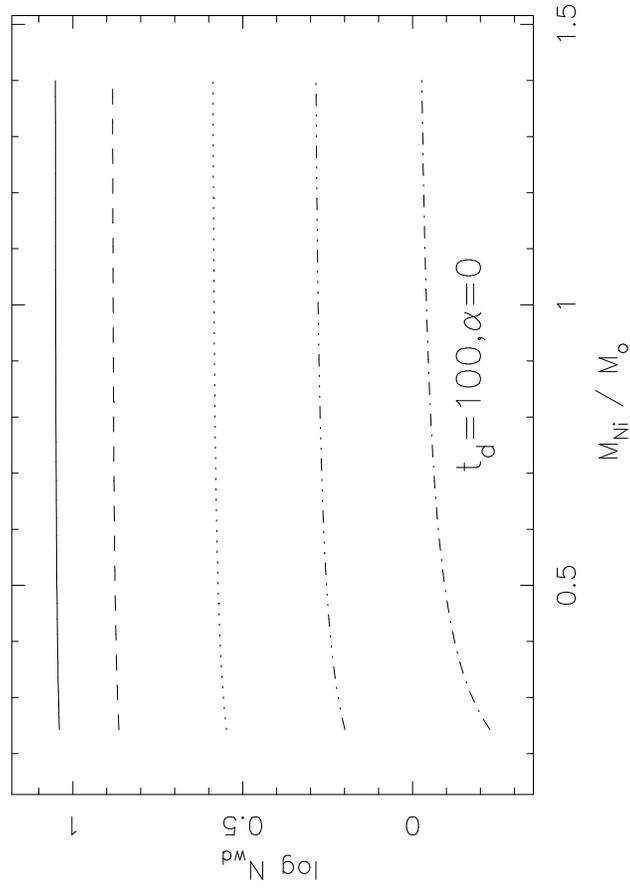
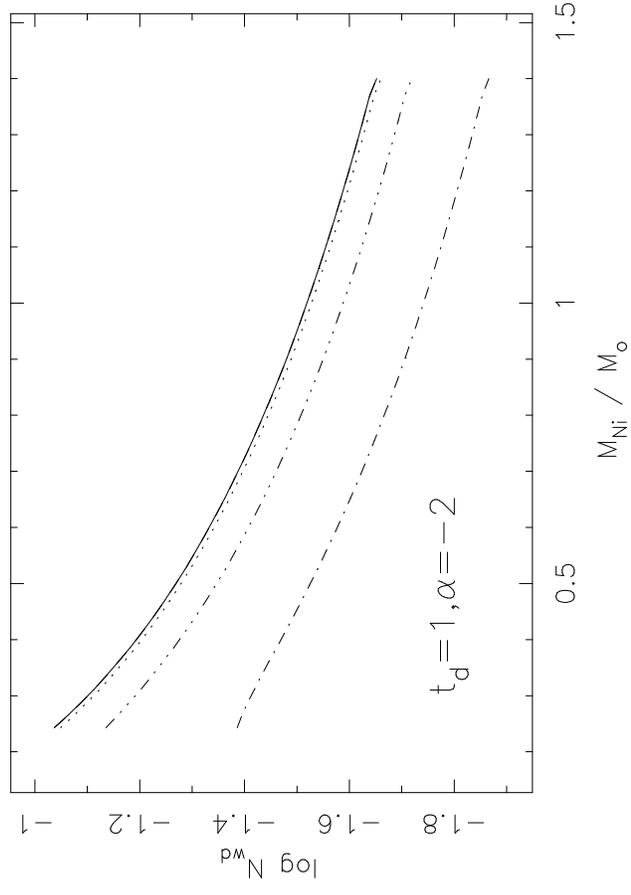
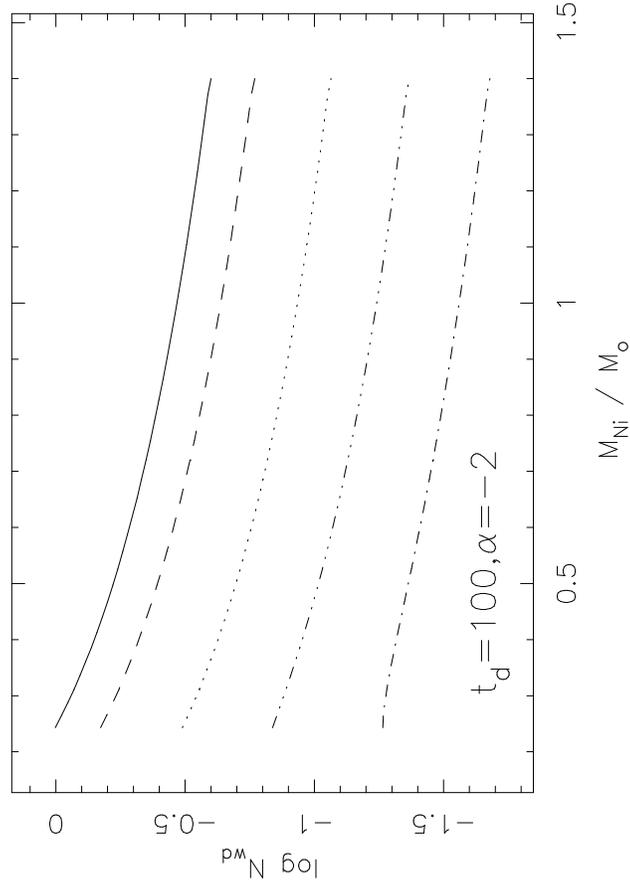
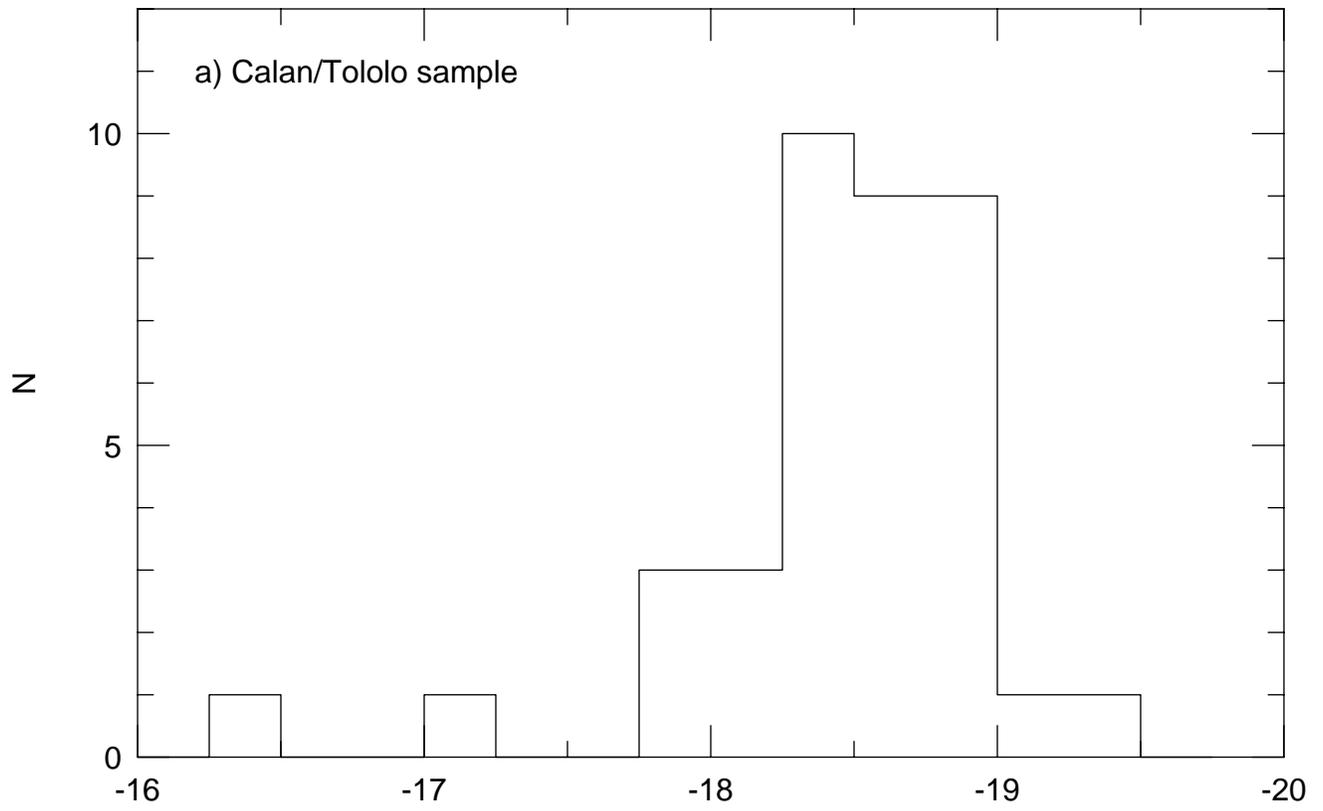
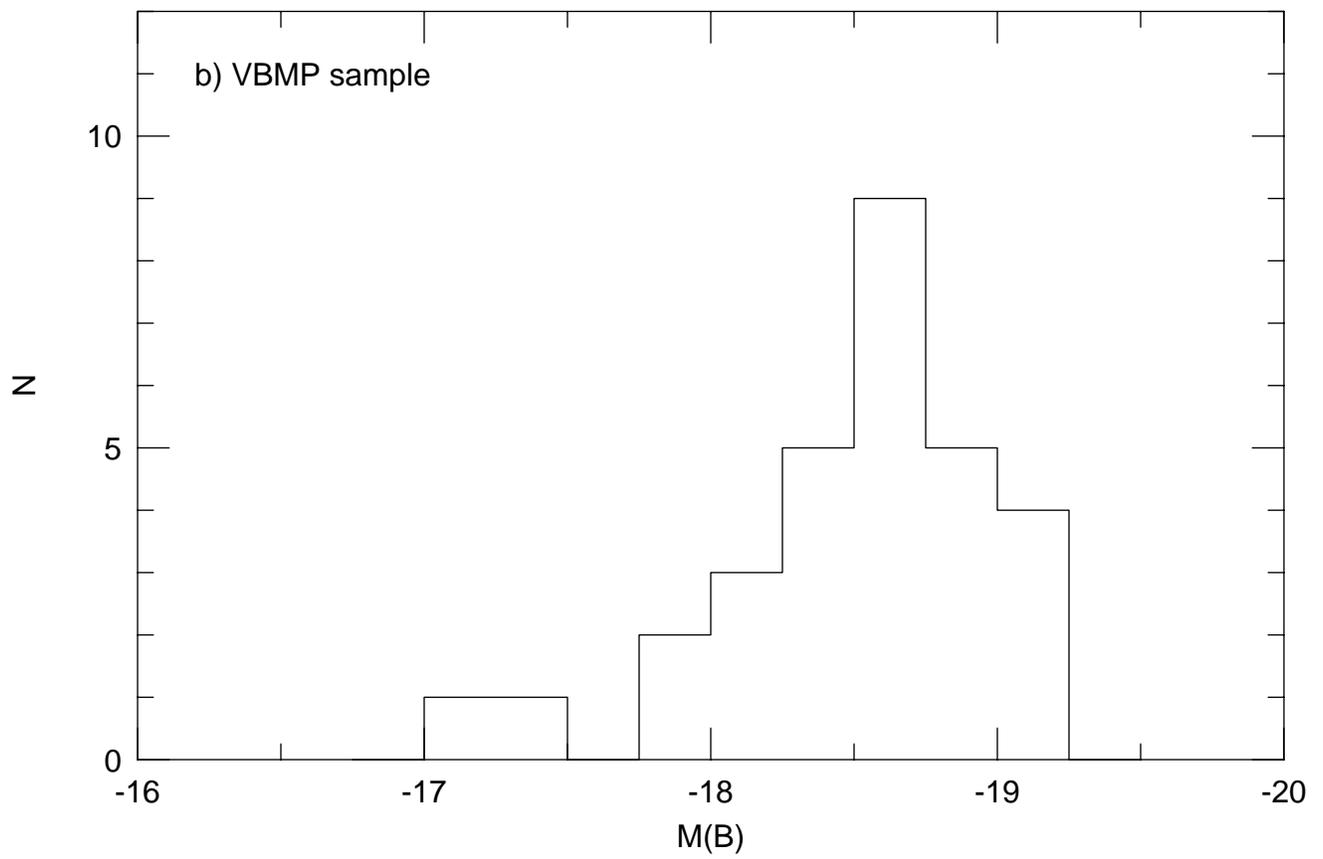
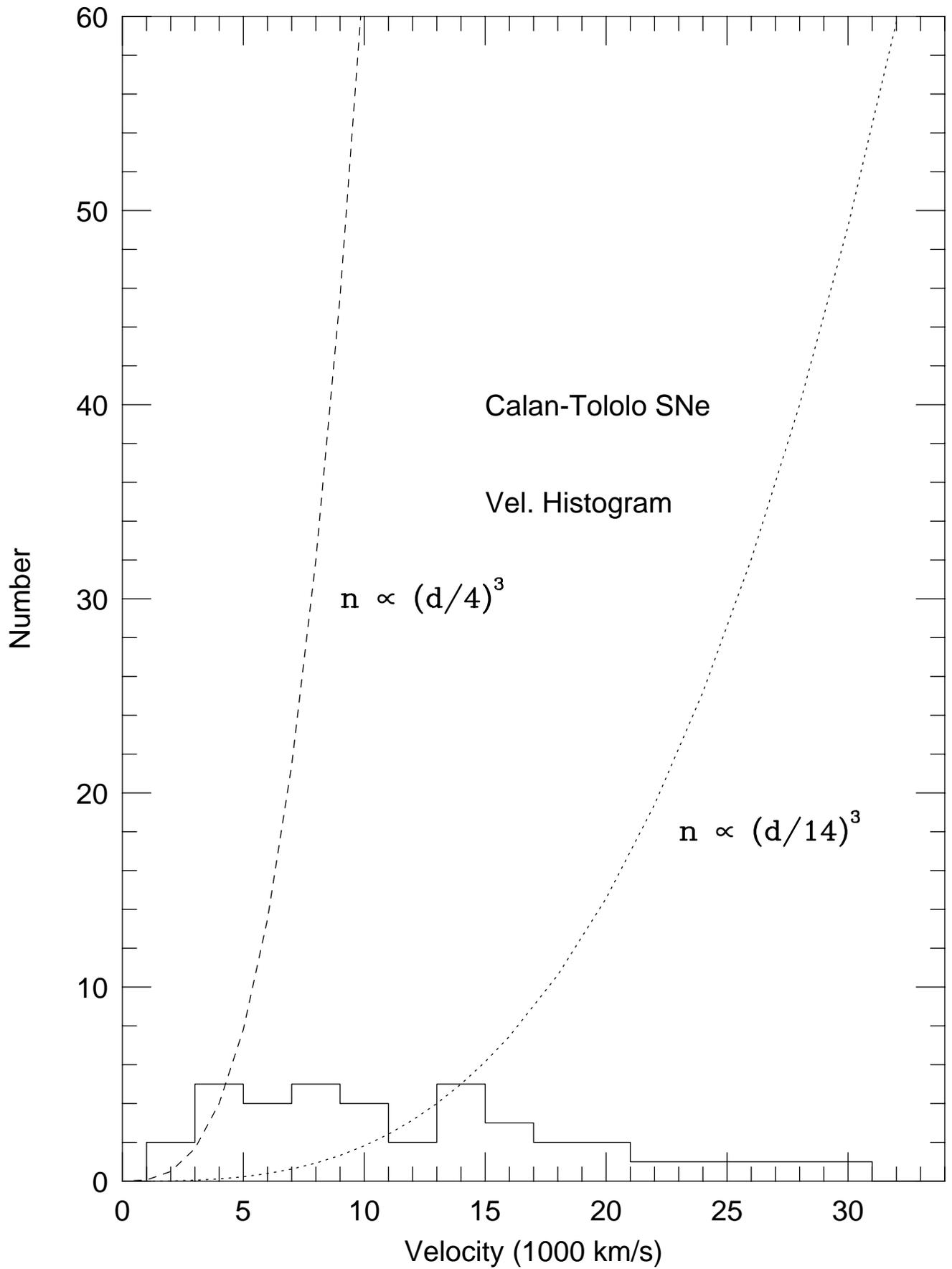
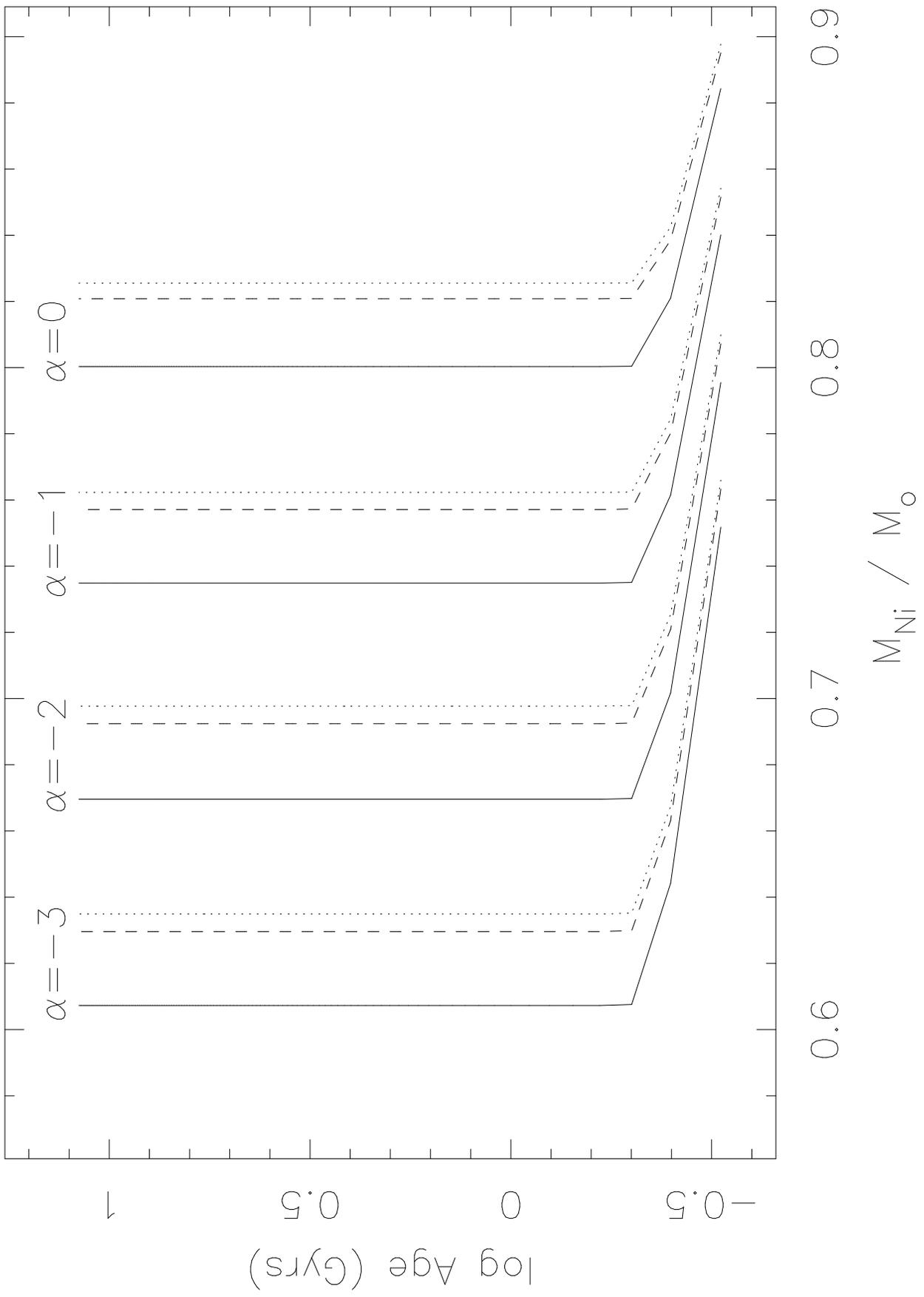
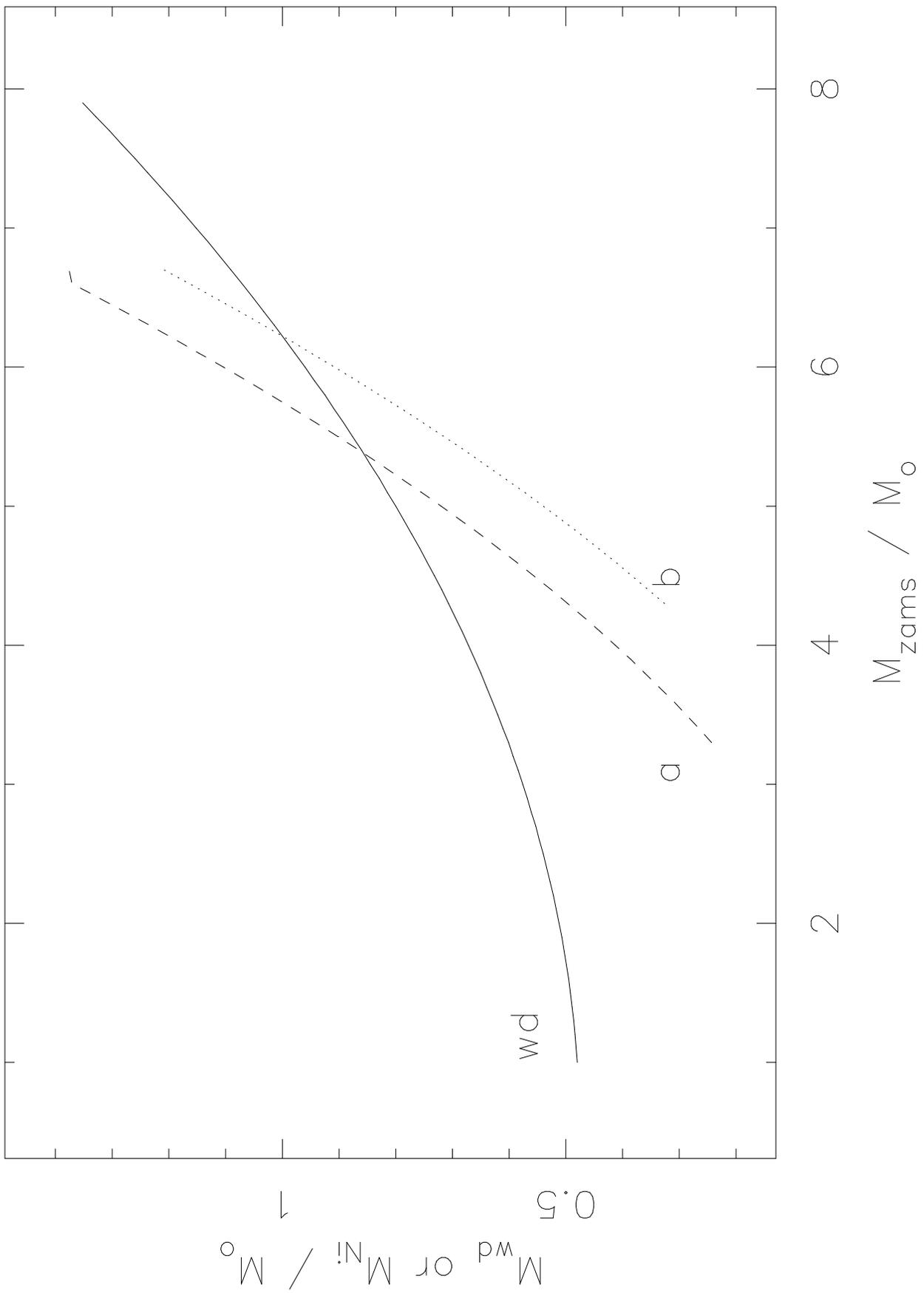
\end{document}